# Tracking Brazil's Real Neutral Rate:

## A Multi-Block Ensemble Framework Combining Statistical Trends, Market Prices, and State-Space Models

*Sample used in this version: 148 monthly observations, February 2014 to May 2026*

*Gabriel de Macedo Santos*

*Instituto de Tecnologia e Liderança*

**Abstract.** This paper presents an implementable framework for tracking Brazil's real neutral-rate proxy, using a block-based ensemble of complementary models. The project begins with daily macro-financial data, converts the series to monthly frequency, computes an ex-ante real Selic rate through the Fisher equation, builds activity-cycle measures from IBC-Br, and then combines five methodological blocks: simple moving averages, statistical trend filters, market-implied curve proxies, a yield-curve state-space model, and a semi-structural IS-Phillips state-space model. The final implementation treats the semi-structural block conservatively: because the IS-Phillips Kalman model falls back to a local-level trend in the current sample, its output is not labeled as structural r-star and receives zero weight in the final ensemble. The latest estimate, for May 2026, places the final operational neutral-rate proxy at 9.48% p.a., with a P25-P75 block range of 8.71%-9.97%. The ex-ante real rate is 10.04%, implying a policy gap of 0.56 p.p. and a neutral stance under the project's thresholds. This neutral classification should be read strictly relative to the project's elevated operational proxy, not relative to conventional long-run structural estimates. The high level of the estimate should not be interpreted as a definitive long-run structural neutral rate: it reflects recent Brazilian real-rate dynamics, market pricing, and trend-based measures in a restrictive cycle. The estimate should be interpreted as a short-to-medium-run shadow neutral-rate proxy under current restrictive monetary and risk-premium conditions, not as a steady-state structural equilibrium rate. The main contribution is therefore methodological and applied: the project offers a transparent, auditable, and extensible measurement system for tracking r-star proxies and monetary-policy stance in Brazil.

# 1. Introduction

The real neutral interest rate is one of the most useful and most difficult objects in monetary-policy analysis. It is useful because it provides a benchmark for assessing whether the ex-ante real policy rate is restrictive, neutral, or expansionary. It is difficult because it is not observed directly, changes over time, and depends on slow-moving forces such as potential growth, risk premia, fiscal conditions, financial integration, demographics, productivity, and inflation credibility.

This project addresses that problem from an applied perspective. Rather than relying on a single fragile estimate of r-star, it builds a measurement system that combines several imperfect but informative signals. The final output is an ensemble estimate that can be monitored over time, decomposed by methodological block, and stress-tested through sensitivity exercises.

The project makes three specific contributions. First, it constructs an auditable, reproducible pipeline for neutral-rate proxies in Brazil. Second, it separates the estimation into statistical, market, state-space, and semi-structural blocks, allowing transparent decomposition. Third, it implements a conservative exclusion rule that removes the semi-structural block when the Kalman model fails stability diagnostics, preventing mislabeled structural output from entering the final estimate.

The empirical object should be interpreted carefully. In the current sample, the project does not deliver a clean structural estimate in the Laubach-Williams sense. The semi-structural IS-Phillips state-space model fails the stability criteria and falls back to a local-level trend of the real rate. The implementation correctly excludes this fallback from the final semi-structural block. Therefore, the final estimate is best described as an operational neutral-rate proxy or dashboard estimate, combining real-rate trends, market information, and a yield-curve state-space signal.

This positioning is important because the latest project estimate, 9.48% p.a. in May 2026, is much higher than longer-run official benchmarks for Brazil published before the latest observations in the sample. That difference is not necessarily an error; it reflects the fact that the project places weight on the recent level and trend of ex-ante real rates and on market-curve proxies. But it does mean that the number should not be presented as a definitive long-run equilibrium rate. The stronger claim is that the framework organizes the evidence transparently and helps separate model signal from model fragility.

## 2. Related Literature

The natural-rate literature begins from the idea that monetary policy should be evaluated relative to a real interest rate consistent with output near potential and inflation near target. Laubach and Williams (2003) formalized this idea in a small semi-structural state-space model in which potential output, trend growth, and the natural rate are jointly estimated with the Kalman filter. Holston, Laubach, and Williams (2017) extended the approach internationally, highlighting that r-star can move persistently and that estimates are surrounded by considerable uncertainty.

The current project follows the spirit of this literature by treating r-star as time-varying and by using state-space methods, but it deliberately avoids overstating what the current data can identify. This conservative interpretation is consistent with the Brazilian evidence in Maka (2023), which emphasizes that estimates of the natural rate are highly sensitive to model specification, initial values, and real-time information constraints.

A second strand of the literature uses statistical filters to separate trend and cycle. The project uses the Hodrick-Prescott filter with the Ravn-Uhlig frequency adjustment for monthly data, Christiano-Fitzgerald band-pass filtering, Butterworth filtering, and local linear trends. These tools are common in macroeconomic applications, but they are not structural models. Hamilton's critique of the HP filter is particularly relevant here: two-sided filters can generate spurious dynamics and are vulnerable to endpoint revisions. The project addresses this by labeling statistical trends explicitly as non-structural and by reporting sensitivity exercises rather than treating one filter as the truth.

A third strand extracts information from the term structure. Market interest rates include expectations about future short rates, risk premia, liquidity premia, and fiscal or country-risk compensation. The project uses curve-based proxies as informative but imperfect market signals. This interpretation is aligned with term-premium models such as Kim and Wright (2005) and Adrian, Crump, and Moench (2013), and with the Brazilian Central Bank's own discussion that neutral-rate estimates should be produced through several methodologies and frequently reassessed.

# 3. Data and Project Architecture

The project starts from daily data and collapses the dataset to monthly frequency using end-of-month observations. The final dataset contains 148 monthly observations, from February 2014 through May 2026. The main inputs include the Selic rate, IPCA 12-month expectations, IBC-Br, DI swap rates, Brazilian government bond yields, U.S. Treasury yields, and Brazil CDS.

| Input block | Variables used | Role in the project |
|---|---|---|
| Policy and inflation | Selic; IPCA 12-month expectation | Computation of the ex-ante real policy rate. |
| Activity | IBC-Br | Construction of activity-cycle and output-gap proxies. |
| Domestic yield curve | Swap DI 3m; Swap DI 10Y; BZ10 bond | Market-implied neutral-rate proxies and yield-curve state-space model. |
| External conditions | U.S. 3m and 10y yields; Brazil CDS | Controls and components for market-based interpretation. |
| Project diagnostics | Kalman status flags; sensitivity outputs; block estimates | Audit trail for model validity and final aggregation. |

*Table 1. Data inputs and economic role in the project.*

*Table 1b. Data source and transformation audit for replication.*

| Variable | Source | Code/ticker | Original frequency | Monthly transformation | Use |
|---|---|---|---|---|---|
| Selic | BCB/SGS or market data provider | Document exact series code/ticker | Daily | End-of-month | Nominal policy rate and Fisher equation |
| IPCA 12m expectation | Focus survey or expectations database | Document exact series code/ticker | Daily/weekly | End-of-month or latest available in month | Expected inflation in Fisher equation |
| IBC-Br | BCB/SGS | Use seasonally adjusted level series when available | Monthly | Level/log transformation; current file treated as growth-cycle proxy | Activity cycle and output-gap proxy |
| Swap DI 3m | Bloomberg | Document exact ticker | Daily | End-of-month | Short end of domestic curve |
| Swap DI 10Y / BZ10 | Bloomberg | Document exact ticker | Daily | End-of-month | Long end of curve and market proxy |

| U.S. yields and Brazil CDS | Bloomberg | Document exact tickers | Daily | End-of-month | External/risk-premium interpretation |
|---|---|---|---|---|---|

A crucial implementation detail is the treatment of IBC-Br. In the current input file, IBC-Br behaves like a growth or variation series rather than a level index: the code detects this and classifies the activity measure as a growth-cycle proxy. As a result, the activity measure is not a strict log deviation from potential output. This weakens the structural interpretation of the IS-Phillips block and is one reason the paper treats the semi-structural output with caution.

# 4. Methodology

The framework is built as a sequential pipeline. Figure 1 summarizes the logic from raw data to final policy-stance classification.

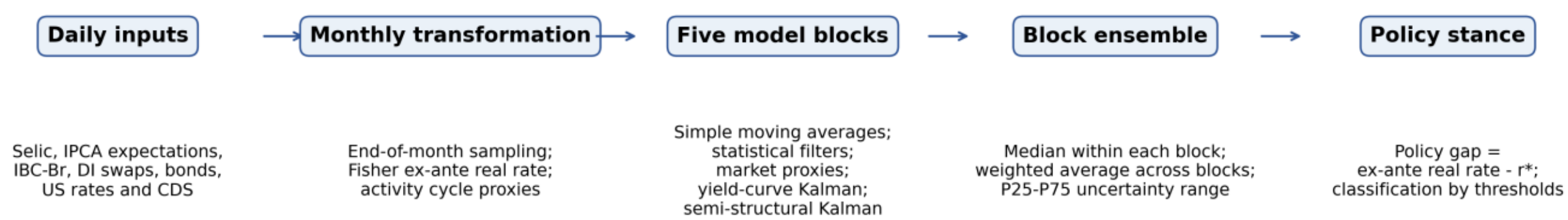

*Figure 1. Project pipeline and economic interpretation of each step.*

## 4.1. Ex-ante real rate

The core observable object is the ex-ante real policy rate. The project computes it using the Fisher equation:

$$r^{\text{treal}} = \left[\frac{(1 + i^{\text{t}})}{(1 + E^{\text{t}}[\pi^{12}{}_{\text{m}}])} - 1\right] \times 100$$

This is the policy-rate benchmark used later to calculate the policy gap. In May 2026, the Selic rate is 14.50% and the 12-month IPCA expectation is 4.05%, implying an ex-ante real rate of 10.04% p.a.

## 4.2. Activity cycle and output-gap proxies

The project estimates three activity-cycle measures: a two-sided HP gap, a rolling HP gap using a 60-month window, and a local-linear-trend measure. Because the IBC-Br input behaves like a growth-cycle series, the code correctly labels the output as a growth_cycle_proxy rather than a level_log_gap. The local-linear activity gap collapses in the sample and is replaced by the rolling HP measure. This behavior is explicitly recorded through output_gap_local_linear_source = hp_rolling_fallback.

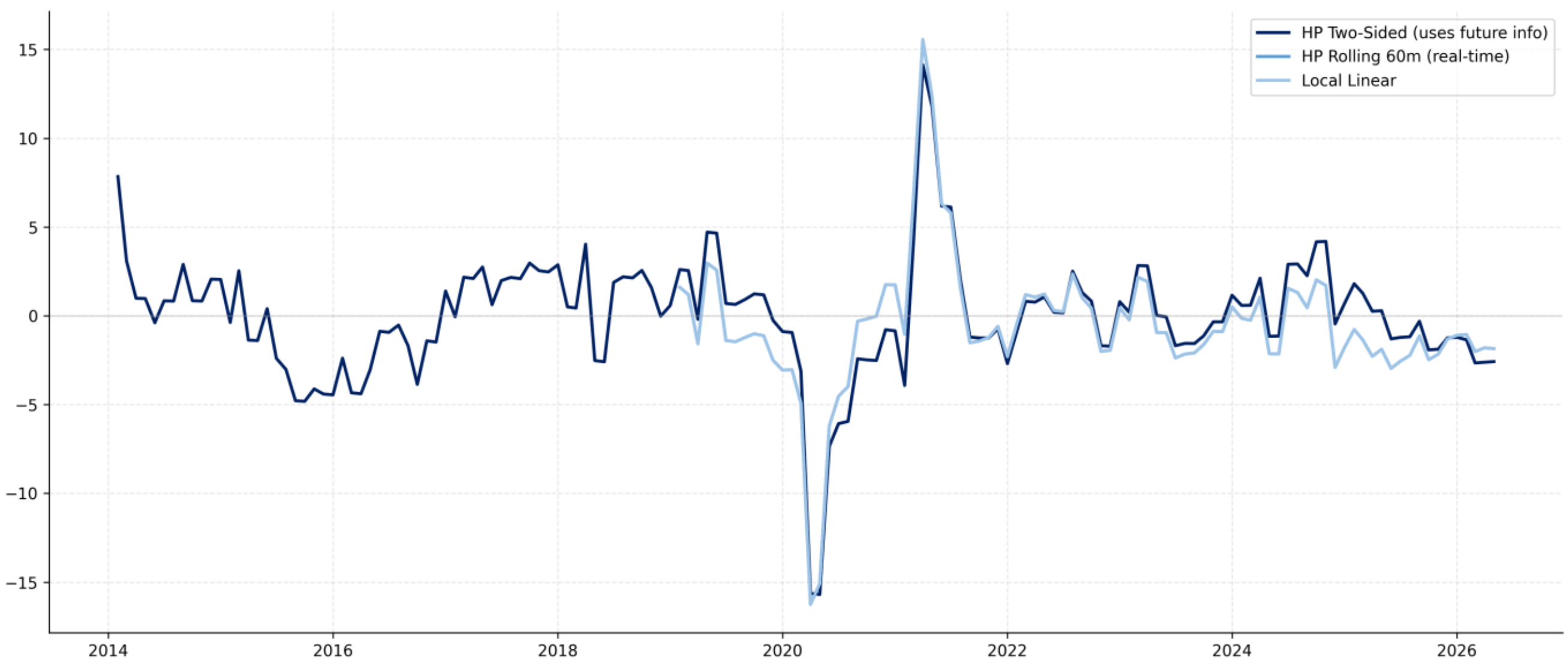

*Figure 2. Activity-cycle measures. The local-linear series is a fallback to the rolling HP gap in the current implementation.*

### 4.3. Statistical trend block

The statistical block applies five trend-extraction methods to the ex-ante real rate: HP, Ravn-Uhlig-adjusted HP, Christiano-Fitzgerald, Butterworth, and local linear trend. The block estimate is the median across available trends. This design reduces the influence of any single filter and recognizes that statistical trends are correlated but not identical.

These estimates are intentionally labeled as statistical trends of the real rate, not structural r-star. That distinction is important because filters can be revised at the end of the sample and can inherit the recent level of the real rate during restrictive cycles.

### 4.4. Market-proxy block

The market block uses the slope between DI 10-year and DI 3-month rates, the bond slope, excess slopes relative to a 2016-2019 baseline, and simplified curve-based neutral-rate proxies. The slope is not treated as a pure term premium. It is a mixture of expected future policy, risk premium, liquidity premium, fiscal premium, and country-risk compensation. The project therefore uses these series as market-implied proxies rather than as a formal arbitrage-free decomposition.

Operationally, the market block should be read as a reduced-form signal: long nominal rates are interpreted as a combination of expected future Selic, inflation compensation, term premium, fiscal and country-risk premia, and liquidity premia. In future versions, this block should be strengthened by separating expected short-rate components from term premia, ideally through an ACM-style term-premium model or a simplified affine decomposition. Until then, these inputs remain market-implied neutral-rate proxies rather than structural estimates.

### 4.5. State-space models

The project implements two state-space components. The first is a yield-curve Kalman model that extracts a smoothed r_star_yield series from short and long rates. This block is retained in the final ensemble as the state-space market block. The second is a semi-structural IS-Phillips model, intended to link activity, inflation expectations, and the real-rate gap. In the current sample, the semi-structural model falls back to a local-level trend of the real rate. The implementation correctly identifies this status as fallback_local_level, avoids labeling r_star_smoothed as structural r-star, and assigns zero weight to the semi-structural block.

A compact representation of the intended semi-structural block is:

$$x_t = \alpha_1 x_t - 1 + \alpha_2 (r_t^{real} - r_t^*) + \epsilon_t^x$$

$$\pi_t = \beta_1 \pi_t - 1 + \beta_2 E_t[pi_t + 12] + \beta_3 x_t + \epsilon_t^\pi$$

$$r_t^* = r_t - 1^* + \epsilon_t^r$$

In the current dataset, this block is not used as a structural estimate because the activity measure is not a clean level-based output gap and the Kalman implementation falls back to a local-level trend. The diagnostic checklist is reported in Appendix Table A3.

### 4.6. Ensemble aggregation and policy gap

The final estimate is constructed in two stages. First, each block is summarized by the median of the estimates inside that block. Second, the block medians are combined through a weighted average. In the current run, the semi-structural block is invalid and receives zero weight. The remaining weights are renormalized across the simple block, the statistical block, the market block, and the yield-curve state-space block.

| Block | Methodological content | Original weight | Effective interpretation in current run | Effective weight (current run) |
|---|---|---|---|---|
| A - Simple | 12m, 24m, and 36m moving averages of the real rate | 10% | Retained as a smooth benchmark. | 13.3% |
| B - Statistical | HP, Ravn-Uhlig, Christiano-Fitzgerald, Butterworth, local linear | 25% | Retained; represents non-structural trends. | 33.3% |
| C - Market | Curve and slope-based proxies | 20% | Retained; represents market-implied signal. | 26.7% |
| D - State-space market | Yield-curve Kalman estimate | 20% | Retained; represents state-space market signal. | 26.7% |
| E - Semi-structural | IS-Phillips Kalman estimate | 25% | Excluded because the model is in fallback_local_level. | 0% |

*Table 2. Model blocks and their treatment in the final ensemble.*

The project defines the policy gap as:

$gap^t = r^{treal} - r*^t$

Positive values indicate restrictive policy relative to the estimated neutral-rate proxy. The classification thresholds are: above 2.0 p.p. restrictive; 0.75 to 2.0 p.p. mildly restrictive; -0.75 to 0.75 p.p. neutral; -2.0 to -0.75 p.p. mildly expansionary; below -2.0 p.p. expansionary.

# 5. Results

## 5.1. Final estimate and block decomposition

The final ensemble tracks the sharp rise in Brazil's ex-ante real rate after 2021, but it smooths the level through the block structure. Figure 3 shows the final estimate with the P25-P75 range across block medians. The latest observation, May 2026, places the final operational neutral-rate proxy ensemble at 9.48% p.a., with a block-interquartile range from 8.71% to 9.97%.

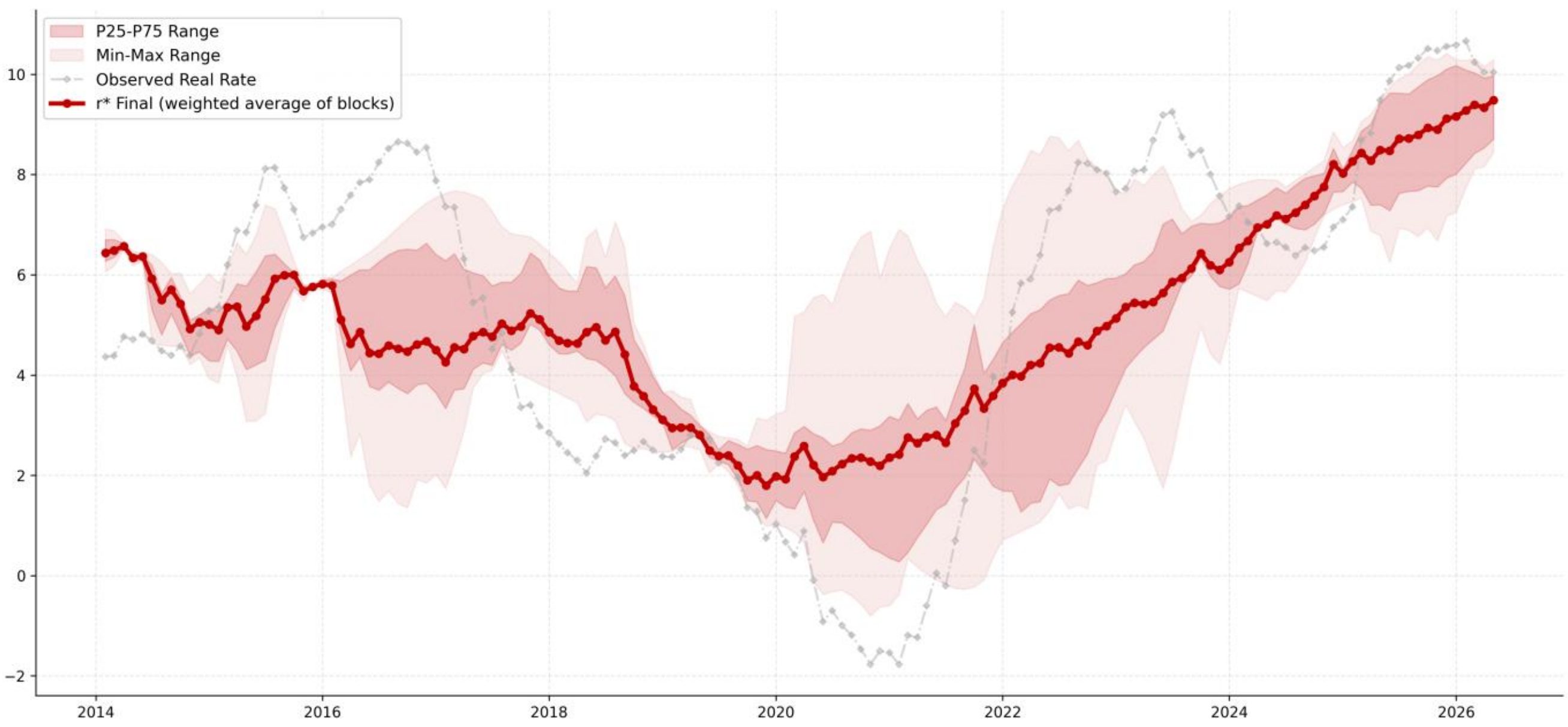

*Figure 3. Final real neutral-rate ensemble and ex-ante real rate.*

Figure 4 summarizes the latest valid block-level estimates used in the ensemble. The simple block is 8.80%, the statistical block is 10.29%, the market block is 8.43%, and the yield-curve state-space block is 9.87%. The final estimate is therefore not driven by a single model; it sits between a high statistical-trend signal and lower market/simple proxies.

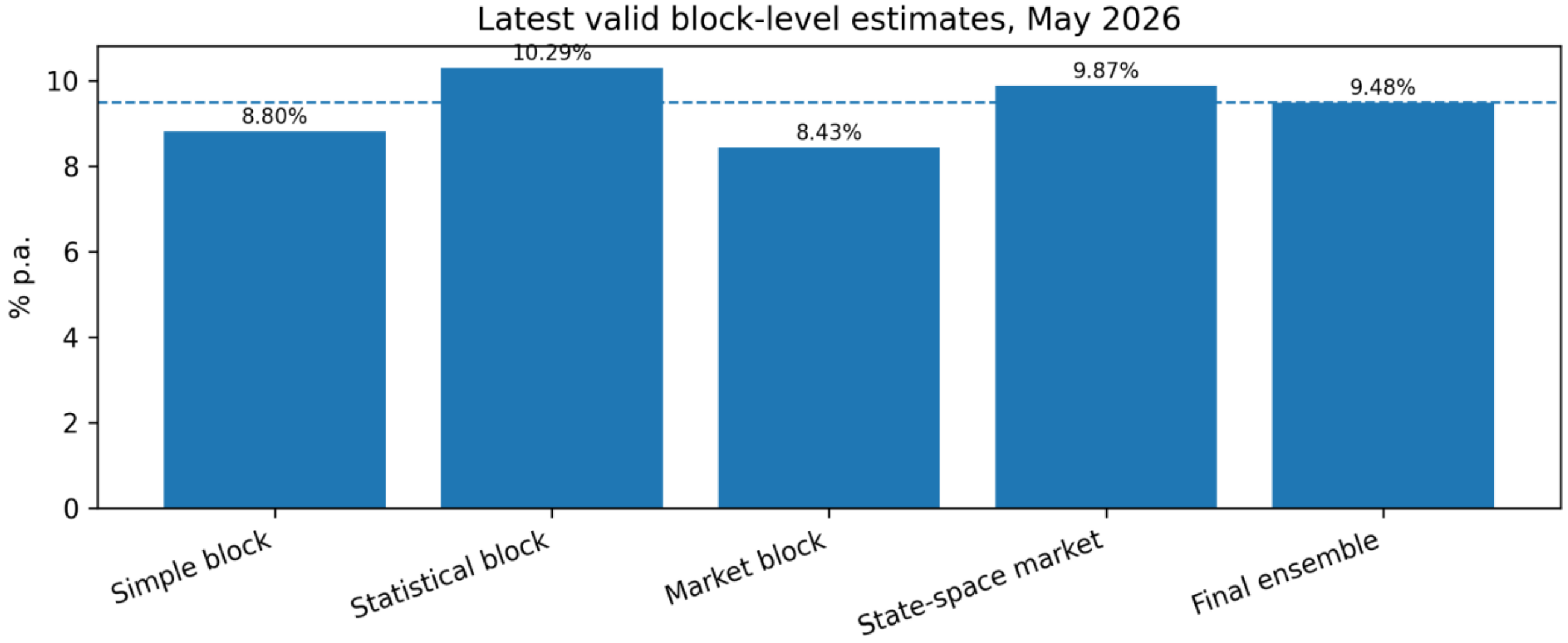

*Figure 4. Latest block-level estimates used in the final ensemble.*

| Metric | May 2026 value | Interpretation |
|---|---|---|
| Selic | 14.50% | Nominal policy rate at end of month. |
| IPCA 12m expectation | 4.05% | Inflation expectation used in the Fisher equation. |
| Ex-ante real rate | 10.04% | Observable real policy-rate benchmark. |
| Final neutral-rate proxy ensemble | 9.48% | Weighted average of valid blocks. |
| P25-P75 range | 8.71%-9.97% | Dispersion across valid block medians. |
| Policy gap | 0.56 p.p. | Neutral stance under project thresholds. |
| Semi-structural status | fallback_local_level | Excluded from final ensemble as structural r-star. |

*Table 3. Latest results from the final project output.*

## 5.2. Policy stance

The policy gap was positive through early 2026, but it declined as the Selic rate moved from 15.00% to 14.50% and as the final ensemble remained high. In January and February 2026, the framework classified policy as mildly restrictive. By April and May 2026, the policy gap had fallen below 0.75 p.p., shifting the classification to neutral.

Important: “neutral” in this framework means neutral relative to the project’s elevated operational proxy. It does not imply neutral relative to conventional long-run structural estimates of the neutral rate, which are materially lower.

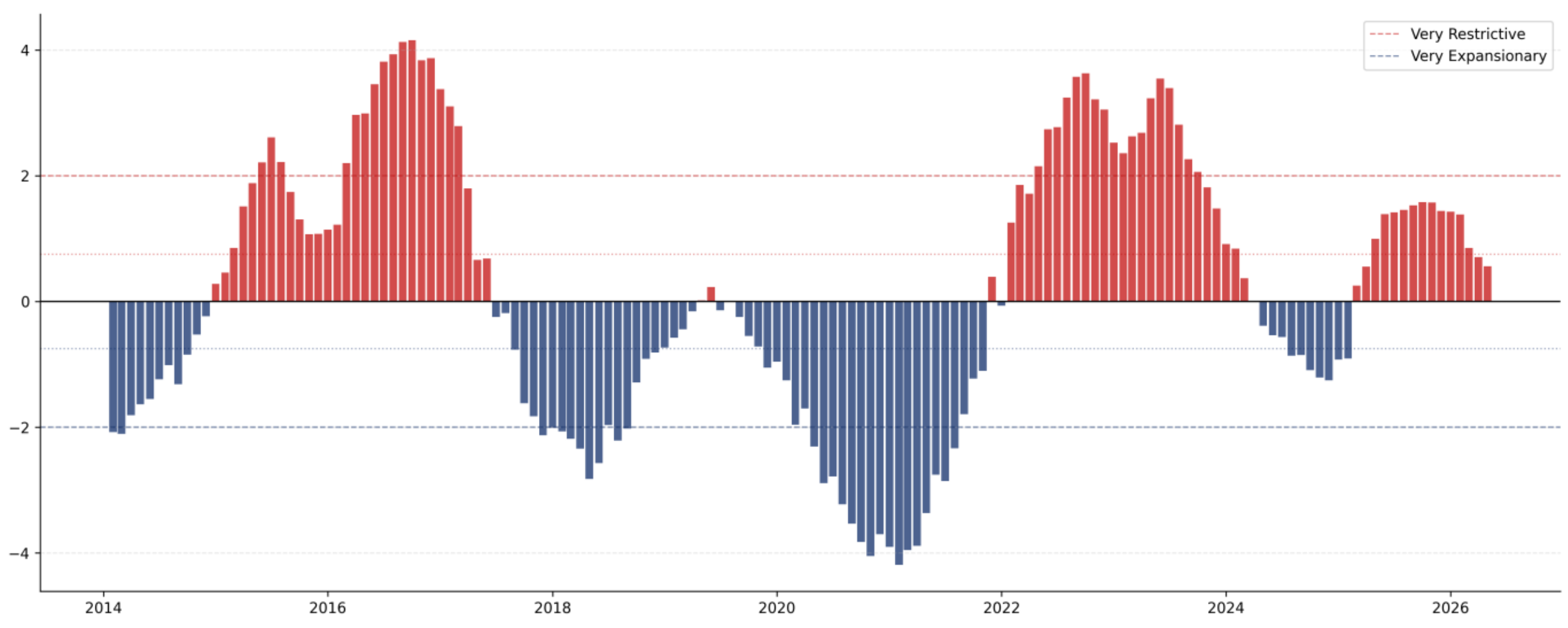

*Figure 5. Policy gap from the final ensemble. Positive values indicate policy above the estimated neutral-rate proxy.*

| Month | Selic | Ex-ante real rate | Final proxy | Policy gap | Stance |
|---|---|---|---|---|---|
| Jan-2026 | 15.00% | 10.59% | 9.16% | 1.43 p.p. | mildly restrictive |
| Feb-2026 | 15.00% | 10.66% | 9.28% | 1.38 p.p. | mildly restrictive |
| Mar-2026 | 14.75% | 10.24% | 9.39% | 0.85 p.p. | mildly restrictive |
| Apr-2026 | 14.50% | 10.04% | 9.34% | 0.70 p.p. | neutral |
| May-2026 | 14.50% | 10.04% | 9.48% | 0.56 p.p. | neutral |

*Table 4. Policy-stance classification during 2026.*

## 5.3. Sensitivity and robustness

The sensitivity exercise confirms that endpoint estimates are model-dependent. The latest HP trend varies from 10.00% to 10.67% depending on the smoothing parameter. Rolling-window alternatives range from 9.73% to 11.10%. Starting the sample in 2014, 2016, or 2018 delivers similar latest HP estimates, all close to 10.7%, but the historical paths differ. This reinforces the paper's main message: the framework is useful as an auditable monitoring system, while any single point estimate should be treated with uncertainty.

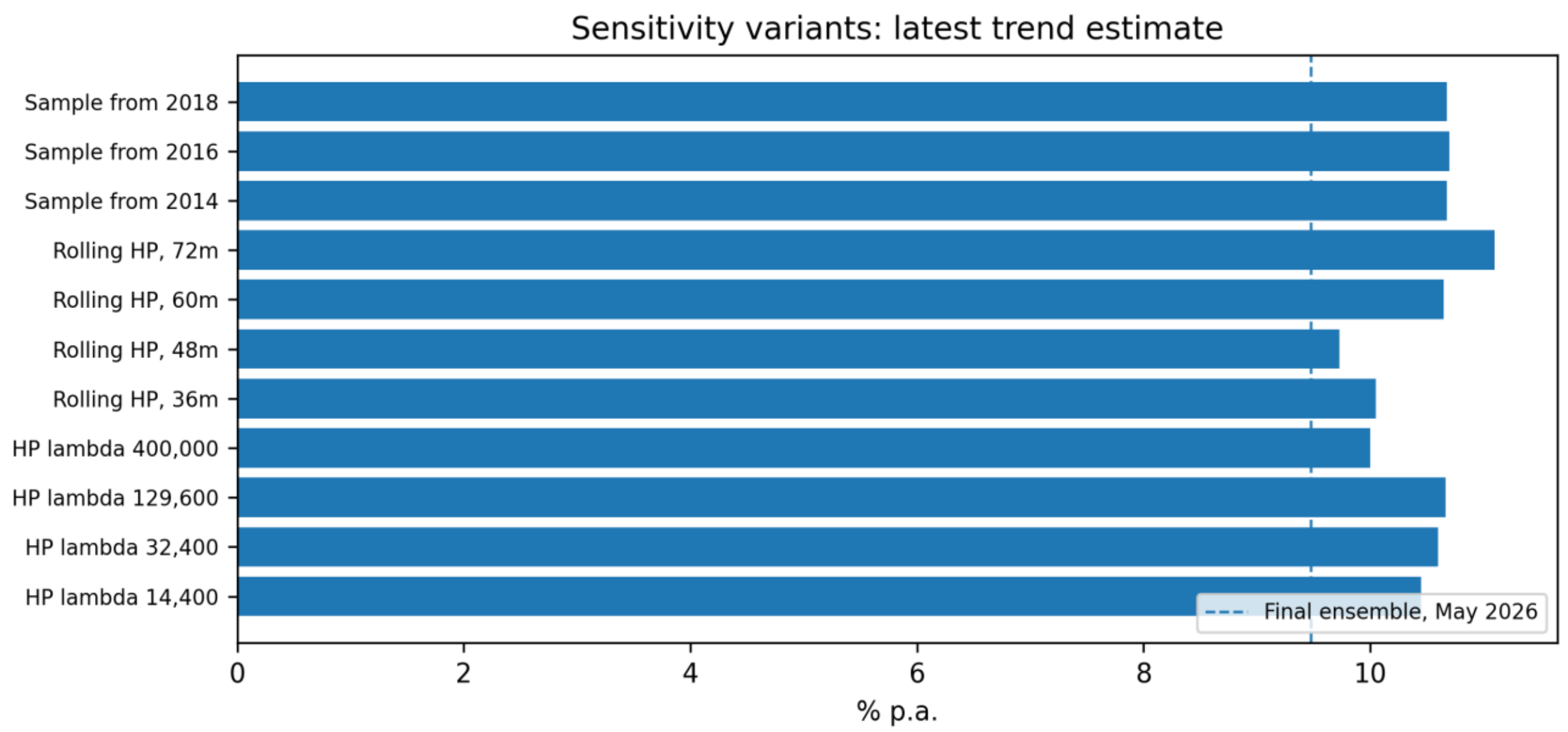

*Figure 6. Sensitivity variants with cleaned labels, compared with the final ensemble.*

| Sensitivity variant | Latest value | What it tests |
|---|---|---|
| HP lambda 14,400 | 10.45% | Lower smoothing parameter. |
| HP lambda 129,600 | 10.67% | Monthly Ravn-Uhlig benchmark. |
| HP lambda 400,000 | 10.00% | Higher smoothing parameter. |
| Rolling HP 48m | 9.73% | Shorter real-time window. |
| Rolling HP 72m | 11.10% | Longer real-time window. |
| Sample from 2018 | 10.68% | Alternative sample start. |

*Table 5. Selected robustness diagnostics from the sensitivity module.*

### 5.4. Comparison with official neutral-rate benchmarks

The Brazilian Central Bank's June 2024 Inflation Report reported an average neutral real interest rate of 4.9%, a median of 5.0%, and a 25th-75th percentile range of 4.7%-5.5% across several methodologies. The project's May 2026 estimate is materially higher. This difference should be interpreted as a methodological and horizon difference rather than as a direct contradiction.

The official benchmark combines survey measures, model-based estimates, market rates discounted by term premia, and external-parity approaches. The current project, by contrast, places substantial weight on recent ex-ante real-rate trends and simplified market-curve proxies during a period in which Brazilian real rates are exceptionally high. Consequently, the final ensemble is closer to a short-to-medium-run operational proxy for the real-rate level consistent with current pricing and recent trends than to a long-run structural equilibrium concept.

*Table 6. Comparison between Central Bank neutral-rate concept and this project.*

| Dimension | BCB (June 2024) | This project (May 2026) |
|---|---|---|
| Concept | Structural neutral / long-run equilibrium | Operational proxy / short-medium run |
| Inputs | Multiple models, surveys, risk-adjusted market data | Real rate trends, market proxies, Kalman yield curve |
| Horizon | Long-run | Short/medium-run |
| Estimate | ~5.0% (median) | 9.48% |
| Interpretation | Structural benchmark for policy calibration | Dashboard estimate for stance monitoring |

## 6. Discussion and Limitations

The first strength of the project is transparency. Every output can be traced to a data transformation, a filter, a state-space estimate, or an explicit aggregation rule. This makes the framework useful for macro research, internal monitoring, and policy-stance communication.

The second strength is conservative labeling. The code does not call a fallback local-level trend a structural r-star estimate. It separates the yield-curve state-space block from the semi-structural IS-Phillips block, excludes the invalid semi-structural block, and reports the fallback status directly. This is exactly the right treatment given the fragility of r-star estimation.

The limitations are equally important. First, the IBC-Br series in the current dataset is treated as a growth-cycle proxy rather than a level index, so the activity gap is weaker as a structural object. Second, statistical filters are subject to endpoint bias and can be heavily influenced by the latest restrictive cycle. Third, market-curve proxies mix expected short rates, risk premia, fiscal risk, liquidity premia, and external conditions. Fourth, the yield-curve state-space model can absorb omitted variables into the latent r-star component. Fifth, the semi-structural model does not converge as a valid IS-Phillips estimate in the current run.

These limitations do not invalidate the project. They clarify its best use. The framework should be presented as a robust monitoring dashboard and ensemble of neutral-rate proxies, not as a final structural estimate. Future versions could strengthen identification by adding realized inflation separately from expectations, using an IBC-Br level index for the output gap, estimating an ACM term-premium model, and applying Bayesian priors to stabilize the semi-structural Kalman model.

## 7. Conclusion

This paper documents a multi-block ensemble framework for tracking Brazil's real neutral-rate proxies. The project integrates daily macro-financial data, monthly transformations, real-rate calculations, activity-cycle measures, statistical filters, market proxies, and state-space models into a single auditable pipeline.

The latest estimate, for May 2026, is 9.48% p.a., with a P25-P75 range of 8.71%-9.97%. Given an ex-ante real rate of 10.04%, the implied policy gap is 0.56 p.p., which the framework classifies as neutral. However, this result should be interpreted in context: the estimate is elevated relative to official long-run benchmarks and reflects the current high-real-rate environment, market pricing, and trend-based methods.

The most credible contribution is methodological. The project shows how to build a disciplined neutral-rate proxy monitoring system that is transparent about which estimates are structural, which are statistical, which are market-implied, and which should be excluded when model diagnostics fail. That discipline makes the framework suitable for applied macro-financial research and provides a clear roadmap for future extensions.

# Appendix A. Implementation audit from the uploaded project

| Audit item | Observed implementation | Assessment |
| --- | --- | --- |
| Semi-structural fallback | kalman_model_status is fallback_local_level for available semi-structural periods. | Correctly excluded from the final ensemble. |
| Block separation | State-space market block uses r_star_yield_smoothed; semi-structural block is separate. | Correct structure for avoiding mislabeled structural output. |
| Activity-cycle source | activity_gap_source is growth_cycle_proxy across the sample. | Important caveat for IS-Phillips interpretation. |
| Local-linear activity gap | output_gap_local_linear_source is hp_rolling_fallback. | Appropriate diagnostic, but reduces structural content. |
| Final output | database_m_ensemble.csv and neutral_rate_estimates.xlsx contain final blocks, ranges, and stance. | Sufficient for article-level analysis and figures. |

*Appendix Table A1. Implementation audit and interpretation.*

*Appendix Table A2. Recommended next-step robustness checks.*

| Check | Purpose | Expected use |
| --- | --- | --- |
| Equal-weight ensemble | Assess dependence on calibrated weights | Benchmark against baseline weights |
| Leave-one-block-out | Identify whether one block dominates the estimate | Report range of final proxy |
| Alternative inflation expectations | Test Focus mean/median and horizon choices | Stress-test ex-ante real rate |
| Different DI maturities | Test 1Y/2Y/5Y/10Y curve choices | Assess market-block sensitivity |
| Level-based IBC-Br gap | Replace growth-cycle proxy with log-level output gap | Re-estimate semi-structural block |

*Appendix Table A3. Semi-structural Kalman diagnostic checklist for the current run.*

| Diagnostic | Current evidence | Assessment |
| --- | --- | --- |
| Final model status | kalman_model_status = fallback_local_level | Fail for structural interpretation |
| Activity-gap input | activity_gap_source = growth_cycle_proxy | Weak structural content |
| Local-linear activity gap | output_gap_local_linear_source = hp_rolling_fallback | Diagnostic fallback |
| Semi-structural block weight | 0% in final ensemble | Appropriate exclusion |
| Structural labeling | r_star_smoothed is not labeled structural r-star | Correct conservative treatment |
| Required future test | Convergence, variance positivity, sign restrictions, and plausible state range | To be reported after re-estimation |